\documentstyle[12pt]{article}
\begin{document}
\pagestyle{empty}
\quad\vskip 1.5cm
\centerline{               \hfill FERMILAB-PUB-96/200-T}
\centerline{               \hfill NHCU--HEP--96--21}
\centerline{hep-ph/9608313 \hfill UICHEP-TH/96--11}
\vskip 1cm
\centerline{\bf \large New Limits on $R$--Parity Breakings}
\centerline{\bf \large in Supersymmetric Standard Models}
\vskip 1cm
\centerline{
Darwin Chang$^{(1,2,3)}$ and
Wai-Yee Keung$^{(2,3)}$ }\vskip 1cm
\centerline{\small\it 
$^{(1)}$Physics Department,
National Tsing-Hua University, Hsinchu, Taiwan, R.O.C.}
\centerline{\small\it 
$^{(2)}$Department of Physics, University of Illinois at Chicago, 
Illinois 60607--7059}
\centerline{\small\it 
$^{(3)}$Fermi National Accelerator Laboratory, P.O. Box 500, Batavia, 
Illinois 60510}
%
\vskip 1.5cm
\begin{abstract}
New limits on couplings $\lambda^i_{jk}{''}$, which
break both the baryon number and the $R$--parity, are derived 
by using a new mechanism that contributes to the 
neutron-anti-neutron oscillation.  The constraints due to 
proton decay and its potential phenomenology are also reexamined.
%
%
\end{abstract}

%
%
\newpage
%
\pagestyle{plain}
\def\lm{\lambda}
\def\lusd{\lm^u_{sd}{''}}
\def\lubd{\lm^u_{bd}{''}}
\def\lubs{\lm^u_{bs}{''}}
\def\lcsd{\lm^c_{sd}{''}}
\def\lcbd{\lm^c_{bd}{''}}
\def\lcbs{\lm^c_{bs}{''}}
\def\ltsd{\lm^t_{sd}{''}}
\def\ltbd{\lm^t_{bd}{''}}
\def\ltbs{\lm^t_{bs}{''}}

The minimal supersymmetric standard model (MSSM)~\cite{one}
has been widely considered as 
a leading candidate for new physics beyond Standard Model.  
However, unlike the Standard Model, additional 
symmetry, called $R$--parity defined as $(-1)^{3B+L+F}$, 
has to be imposed on the minimal supersymmetric extensions of 
Standard Model (MSSM) in order to avoid  renormalizable interactions
which violate the lepton and baryon numbers.
It is in fact one of the main theoretical weakness of these models because  
the conservation of $R$-parity is an ad hoc imposition without 
fundamental theoretical basis.  
Therefore, it is interesting to ask how small these 
R-parity breaking couplings have to be by investigating the 
phenomenological constraints on them 
if they are indeed added to the MSSM~\cite{two}.

The most general renormalizable $R$-violating superpotential using only 
the MSSM superfields is 
\begin{equation}
W=\lm^k_{ij}L_iL_j\bar{E}_k+\lm_{ijk}'L_iQ_j\bar{D}_k
+\lm^i_{jk}{''}\bar{U}_i\bar{D}_j\bar{D}_k \ . 
\end{equation}
Here, $i,j,k$ are generation indices and we have rotated away a
term of the form $\mu_{ij}L_iH_j$.  
The couplings $\lm^k_{ij}$ must be antisymmetric in flavor, 
$\lm^k_{ij}=-\lm^k_{ji}$.  
Similarly, $\lm^i_{jk}{''}=-\lm^i_{kj}{''}$.  
There are 36 lepton number non-conserving couplings (9
of the $\lm$ type and 27 of the $\lm'$ type) and 9 baryon number 
non-conserving couplings (all of the $\lambda{''}$ type) in Eq.(1).

To avoid rapid proton decay, it is usually assumed in the literature that 
$\lm$, $\lm'$ type couplings do not coexist with 
$\lambda{''}$ type couplings.
We  make the same assumption here and consider constraints 
on $\lambda{''}$ couplings only.  The constraints on $\lm$ and $\lm'$
couplings~\cite{three} have been discussed quite extensively in the
literature.  
We  discuss a new set of constraints on $\lambda{''}$ 
due to their contributions to the neutron-anti-neutron oscillation (NANO)
through a new mechanism.  
In addition, we wish to emphasize that 
couplings of $\lambda{''}$ type cannot give rise to the proton decay only when
the proton is the lightest particle with 
$(-)^L = +$ and $(-)^F = -$.   If there are lighter fermions with 
$(-)^L = +$, such as lightest neutralino or gravitino,
the proton can in principle decay into them.  In that case strong constraint
on $\lambda{''}$ can be derived and, if they indeed give the leading 
contributions to the proton decay, it will greatly affect the leading
proton decay mode.

There are many existing constraints on the 
9 different $\lambda{''}$ couplings, $\ltbs, \ltbd,$
$\ltsd,$ $\lcbs,$ $\lcbd, \lcsd, \lubs, \lubd$ and $\lusd$.  
First of all, one can show 
~\cite{br,goity} that the requirement of perturbative 
unification typically places a bound of order one  
on many of the couplings.  
A potential constraint on $|\lusd|$ due to neutron-anti-neutron 
oscillation was discussed in ref.~\cite{zw}.  
However it was later realized~\cite{goity,barbieri} that the 
constraint is not as strong as it was originally derived 
due to additional source 
of suppression factors.  
Stronger constraints~\cite{goity,zw,barbieri} on 
$|\lusd|$ can be derived from the non-observation of
double nucleon decay into two kaons (such as
${}^{16}O\rightarrow {}^{14}C\ K^+ K^+$),
\begin{equation}
|\lusd| < 10^{-6} \quad  
({\rm for\  } M_{\tilde q} \simeq 300 {\rm \ GeV})
\ .
\end{equation}
In addition, Goity and Sher~\cite{goity} was able to find a one--loop 
mechanism which gives rise to a strong bound on $|\lubd|$ 
by the non-observation of neutron-anti-neutron oscillations. 
For squark masses of 300 GeV, their bound is roughly  
\begin{equation}
|\lubd| < 5\times 10^{-3} \ .
\end{equation}
Both numerical values have large uncertainty due to the nuclear matrix
elements.  In these references, some bounds on products of couplings
were obtained by considering $K$-$\bar{K}$ mixing.  Recently, bounds
from the $\bar{b}b$ induced vertex correction to the decay of the $Z$
into two charged leptons have been obtained~\cite{cern}; though
potentially interesting, with present data they are not significantly
better than the bound from perturbative unification.  In
Ref.\cite{sher}, Carlson, Roy and Sher obtained some new
bounds on the $\lambda{''}$ couplings from the rare two-body
nonleptonic decays of $B$ and $D$ mesons.  From the recent
experimental upper bound~\cite{cleo} of $5\times 10^{-5}$ on the
branching ratio of $B^+\rightarrow \bar{K}^oK^+$, they obtained
\begin{equation}
|\lm^q_{bs}{''}\lm^q_{sd}{''}| < 5 \times 10^{-3}\tilde{r}_q^2 \quad ,
\quad\hbox{for  } q=t,c,u.
\end{equation}  
where $\tilde{r}_q = m_{\tilde{q}}/m_W$.
For the decay $B^+\rightarrow \bar{K}^o\pi^+$ or $B^-\rightarrow
{K}^o\pi^-$, using the experimental upper bound~\cite{cleo} of 
$5\times 10^{-5}$ on ${\rm B.R.}(B^+\rightarrow\bar{K}^o\pi^+)$,
they obtained 
\begin{equation}
|\lm^q_{bd}{''}\lm^q_{sd}{''}| < 4.1\times 10^{-3}\tilde{r}_q^2 \quad , 
\quad\hbox{for  } q=t,c,u.
\end{equation}
Barbieri and Masiero~\cite{barbieri} also obtained bounds on $\lambda{''}$
from their one--loop contributions 
to the $K_L$-$K_S$ mass difference.  
These bounds can be summarized as~\cite{goity} 
\begin{eqnarray}
|\ltbs\ltbd| &<& \hbox{\rm min}\left(6\times 10^{-4}\tilde{r}_t,
3\times 10^{-4}\tilde{r}_t^2\right) \quad ,        \nonumber\\
|\lcbs\lcbd| &<& \hbox{\rm min}\left(6\times 10^{-4}\tilde{r}_c, 
2\times 10^{-4}\tilde{r}_c^2\right) \quad ,
\end{eqnarray}
assuming that the top quark is much lighter than
scalar top, and that all squark masses are degenerate.  
Bounds from $D$-$\bar{D}$ mixing also have been considered~\cite{goity}
and give 
\begin{equation}
|\lcbs\lubs| < 3.1\times 10^{-3}\tilde{r}_s \quad.
\end{equation}

Let us first consider the constraints on $\lambda{''}$ imposed 
by the non-observation of NANO. 
First of all, it is clear that in order to violate baryon number 
by two units one needs to use $\lambda{''}$ twice.
Goity and Sher~\cite{goity} discovered a one--loop diagram 
as in Fig.~1 that can contribute to NANO.
The resulting effective operator is~\cite{goity}
	\begin{equation}
{\cal L}^{(1)}_{\hbox{\tiny BX}}=
T_1 \epsilon_{\alpha\beta\gamma}\epsilon_{\alpha'\beta'\gamma'}
\bar{u}^c_{R\alpha} 
d_{R\beta} 
\bar{d}^c_{L\gamma} 
d_{L\gamma'} 
\bar{u}^c_{R\alpha'} 
d_{R\beta'} 
\quad .
      \end{equation}
The  diagram is calculated at zero external momenta and yields
\def\Mu{{\bf\mu}}
\begin{equation}
T_1=-{g^4 \over 32\pi^2}
\lambda^u_{dk}{''} \lambda^u_{dj}{''} 
U_{nd}U^*_{nj}U_{id}U^*_{ik}
m_{\tilde w}
{
\Mu_{\tilde{d_k}}^2 \Mu_{\tilde{d_j}}^2 
\over
M_{\tilde{d_k}}^4   M_{\tilde{d_j}}^4
}
J(M^2_{\tilde{w}},M^2_W,m^2_{u_i},M^2_{\tilde{u}_{n}})
\quad.
\end{equation}
Here we assume the Kobayashi-Maskawa matrix of the
left-handed squark  is
the same as that of the quarks.  
Dummies $i$, $j$, $k$ and $n$ are generation indices
which are summed in (8). 
Note that $j$ and $k$ cannot be $d$.  We follow the convention and Feynman
rule in Ref.\cite{brkp}.
The momentum integral $J$ is
\begin{equation}
J(a_1,a_2,a_3,a_4)
=\int_0^\infty {x^2dx\over \prod_{k=1}^4   (x+a_k)}
=\sum_{i=1}^4{a^2_i\ln(a_i)\over
                           \prod_{k\ne i}(a_i-a_k)}\quad .
\end{equation}
The mass--squared term $\Mu_{\tilde{q}}^2$ which mixes $\tilde q_L$ and
$\tilde q_R$ is given by 
\begin{equation}
\Mu_{\tilde{q}}^2=A\,m_{q} \quad .
\end{equation}
Coefficient $A$ is a soft supersymmetry breaking parameter\cite{goity}. 
Consistent with most of the MSSM in the literature, we assume that all
left--right squark mixing parameters are flavor diagonal.

Phenomenologically, the neutron oscillation time is given by
$\tau=1/\Gamma$, where the transition probability (per unit time)
$\Gamma=|T\psi(0)^2|$.  The amplitude $T$ due to the Feynman diagram in
Fig.~1 is given by $T_1$ in (8).  The wave function squared
$\psi(0)^2$, which is simply related to the matrix element of the
operator in (7), has been estimated by Pasupathy\cite{pasu} to be
$\psi(0)^2= 3\times 10^{-4}$ GeV${}^6$. Nevertheless one should be
aware that other evaluations\cite{pasuothers} differ by more than an
order of magnitude.  Those differences among various 
evaluations characterize the degree of our
ignorance about the matrix element; however, $\lambda^u_{db}{''}$ will
vary only as the square root of $\psi(0)^2$.
From the experimental limit on the 
neutron oscillation time\cite{pdg}, $\tau > 1.2\times 10^8$~s, the
bound on $\lambda^u_{db}{''}$ can be obtained.  

The results depend on
the the squark masses as well as Kobayashi-Maskawa angles.
We shall assumed, as is Ref.\cite{goity}, that the charm and up squark
masses are degenerate.
Since $T_1$ has two powers of the masses of $m_{d_k}$ and $m_{d_j}$,
Goity and Sher assumed that the $b$ squark dominates.  They obtained a
strong bound on $|\lubd|$ of roughly $10^{-3}$ for $M_{\tilde{u}}=
M_{\tilde{c}} = 200$ GeV and $M_{\tilde{t}} \sim $ 220 GeV, for the
scenario $A=M_{\tilde w}=200$ GeV.
Note that if all three up--type squarks have the same mass, the
transition amplitude in (8) vanishes because of the GIM cancelation
via the internal up--type squark line by suppression factors of the
form $\Delta M^2_{\tilde{q}}/M^2_{\tilde{q}}$ where $\Delta
M^2_{\tilde{q}}$ is a typical up--squark mass difference.

This set of diagrams turns out to be just one of the four sets of 
one--loop diagrams that can contribute to NANO.  The other three diagrams
are given in Fig.~2,3,4.  It is not too hard to see that the
contributions from diagrams in Fig.~3 and Fig.~4 are proportional the
external quark momenta and therefore their contributions are
suppressed by additional factor of $m_N/M_W$ where $m_N$ is the neutron
mass.  Since the remaining factors are roughly of the same order of
magnitude, we shall ignore contributions from Fig.~3 and Fig.~4
even though they can be just as easily estimated.

The resulting effective operator from the contribution of Fig.~2 is
	\begin{equation}
{\cal L}^{(2)}_{\hbox{\tiny BX}}=
T_2 \epsilon_{\alpha\beta\gamma}\epsilon_{\alpha'\beta'\gamma'}
\bar{u}^c_{L\alpha} 
d_{R\beta} 
\bar{d}^c_{L\gamma} 
d_{L\gamma'} 
\bar{u}^c_{L\alpha'} 
d_{R\beta'} 
\quad,	\end{equation}
where $T_2$ is estimated to be 
\begin{eqnarray}
T_2&=&-{g^4 \over 16\pi^2}
\lambda^i_{dk}{''}\lambda^n_{dj}{''}
U_{id} U_{uk}^*   
U_{nd} U_{uj}^*
      m_{\tilde{w}}         m_{u_i} m_{d_j}
    \Mu^2_{\tilde d_k}    \Mu^2_{\tilde u_n}
\nonumber\\
&\times& I(M^2_{\tilde w  },M^2_W,m^2_{u_i},m^2_{d_j} ;
          M^2_{\tilde u_n},M^2_{\tilde d_k}) \ .
\end{eqnarray}
Here the function $I(a_1,a_2,a_3,a_4;a_5,a_6)$ of the momentum integral is
defined to be
\begin{equation}
\int_0^\infty {xdx\over(x+a_5)^2(x+a_6)^2 \prod_{k=1}^4 (x+a_k)}
=
{\partial^2\over\partial a_5\partial a_6}
\sum_{i=1}^6{a_i\ln(a_i)\over
\prod_{k\ne i}(a_k-a_i)}
\ .
\end{equation}
The two diagrams Fig.~1 and 2 can give rise to quite different
constraints on $\lambda{''}$.  First of all, the two diagrams involve
quite different operators and therefore their matrix elements may be quite
different also.  For numerical illustration, we shall take the two matrix
elements to be the same.  Secondly, the two contributions, $T_1$ and
$T_2$, involve quite different dependence on $\lm''$ parameters and
various masses.  
Unlike in the case of Fig.~1, the GIM cancelation in case of
degenerate squark masses does not occur in Fig.~2 because the
cancelation is already broken by the generation dependence in the
couplings of $\lambda{''}$.
Using the known quark mixing angles, we found numerically that the
channels for $i$=$n$=$t$ and $k,j$=$s$ or $b$ dominate if all squarks
have the same mass.
Barring from accidental cancelation due to different contributions for 
$k,j=s$ or $b$,  
we obtain the constraints for the scenario $M_{\tilde q}=A=M_{\tilde w}$,
\begin{eqnarray}
  |\lambda^t_{sd}|^2 &<& \left( { 200 {\rm\ MeV}\over m_s} \right)^2
  4.5\times 10^{-6} \quad {\rm for }\quad M_{\tilde q} = 100 {\rm\  GeV}
\ ,\nonumber\\
  |\lambda^t_{sd}|^2 &<& \left( { 200 {\rm\ MeV}\over m_s} \right)^2
  2.4\times 10^{-4} \quad {\rm for }\quad M_{\tilde q} = 200 {\rm\  GeV}  
\ ,\nonumber\\ 
  |\lambda^t_{bd}|^2 &<& 7  \times 10^{-6} 
\quad {\rm for }\quad M_{\tilde q} = 100 {\rm\  GeV}  \ ,\nonumber\\
  |\lambda^t_{bd}|^2 &<& 4 \times 10^{-4} 
\quad {\rm for }\quad M_{\tilde q} = 200 {\rm\  GeV} \ .
\end{eqnarray}

Next we  discuss the issue of the proton decay when the $R$-parity
breaking terms such as $\lambda{''}$ and a light neutralino coexist.  
This possibility was mentioned only briefly in the literature.
As emphasized earlier, if the proton is not the lightest fermion with zero
lepton number, then in general the $\lambda{''}$
coupling will induce the proton to decay into such a lightest supersymmetric
particle (LSP).  For example, if the LSP is a photino
and $m_{\tilde{\gamma}} \ll m_p - m_K$, 
the leading proton decay mode can be be $p^+ \rightarrow K^+
\tilde{\gamma}$ due to the tree-level diagram in Fig.~5.  
Previous search on proton decay mode $p \rightarrow K \nu$ \cite{pdg} can
be translated into the experimental limit on this mode and 
places a very stringent constraint on the coupling 
\begin{equation}
\lm^u_{ds}{''} < 10^{-15} \ , \quad{\rm if}\quad 
m_{\tilde{\gamma}} \ll m_p - m_K \ .
\end{equation}
For a slightly heavier photino, which is still lighter than the proton
$ m_p > m_{\tilde{\gamma}} 
\lower2mm\hbox{$\,\stackrel{\textstyle >}{\sim}\, $}   m_p - m_K$, 
the proton can still decay 
through $p^+ \rightarrow \pi^+ \tilde{\gamma}$ 
or      $p^+ \rightarrow e^+ \nu \tilde{\gamma}$ 
with a weaker bound on $\lambda{''}^u_{ds}$ because  
additional vertices of the 
weak interaction are  needed in the process.  
Also in this case, one can  consider a
tree-level process that  converts two nucleons in a nuclei into
a photino in $NN \rightarrow \Lambda \tilde{\gamma}$.
If the LSP is Higgsino or zino, the limit will be only slightly weakened.

\noindent
{\bf Acknowledgments}
We wish to thank Marc Sher and Rabi Mohapatra for discussions.
This work was supported in parts by National Science Council of
Taiwan, R.O.C. and by U.S. Department of Energy (Grant No. 
DE-FG02-84ER40173).
\newpage
\def\prd#1#2#3{{\rm Phys. ~Rev. ~}{\bf D#1} (19#2) #3}
\def\plb#1#2#3{{\rm Phys. ~Lett. ~}{\bf B#1} (19#2) #3}
\def\npb#1#2#3{{\rm Nucl. ~Phys. ~}{\bf B#1} (19#2) #3}
\def\prl#1#2#3{{\rm Phys. ~Rev. ~Lett. ~}{\bf #1} (19#2) #3}

\bibliographystyle{unsrt}

\section*{Figure Captions}
\begin{enumerate}

\item \label{fig1} 
A one--loop diagram for the amplitude $T_1$ 
in the process $n$--$\bar n$ oscillation.
The $\lambda{''}$ couplings appear in the circles.
\item \label{fig2} 
A one--loop diagram for the amplitude $T_2$ 
in the process $n$--$\bar n$ oscillation.
\item \label{fig3} 
A one--loop diagram for the amplitude $T_3$ 
in the process $n$--$\bar n$ oscillation.
This amplitude is suppressed by $m_N/M_W$.
\item \label{fig4} 
A one--loop diagram for the amplitude $T_4$ 
in the process $n$--$\bar n$ oscillation.
This amplitude is suppressed by $m_N/M_W$.
\item \label{fig5} 
A tree--level diagram for the proton decay 
$p^+\rightarrow K^+\tilde\gamma$
for a very light photino.
\end{enumerate}
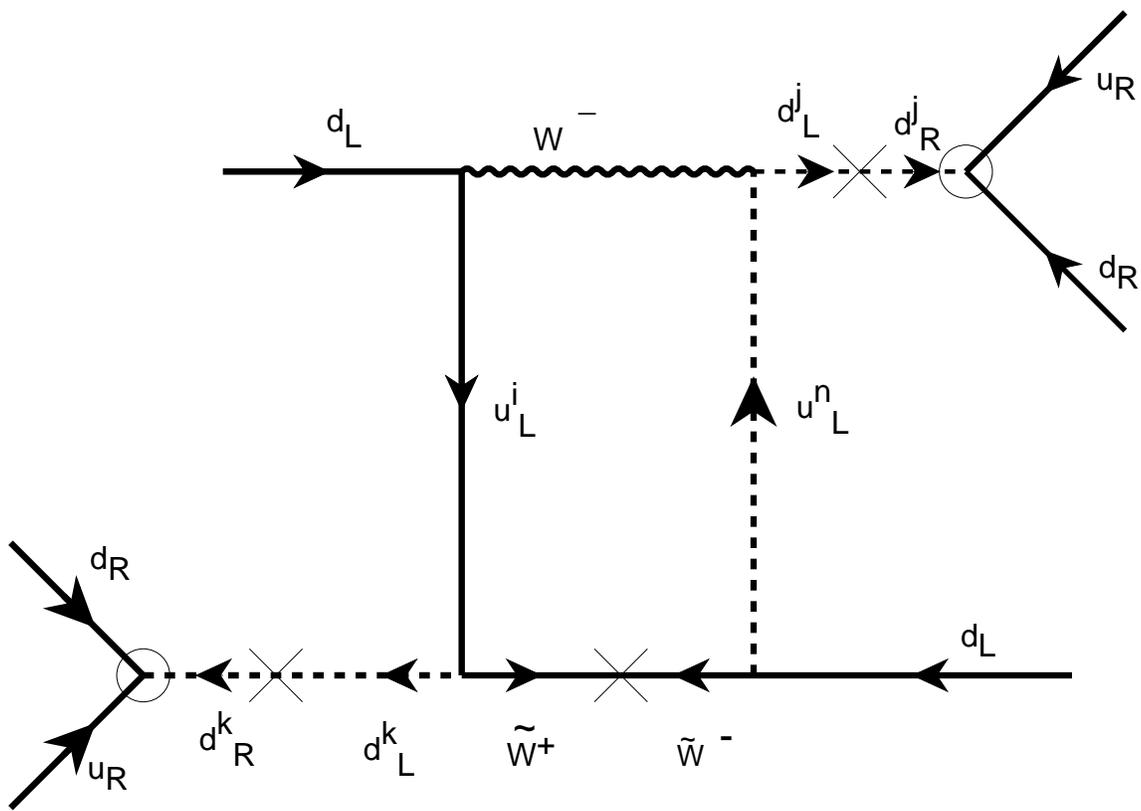
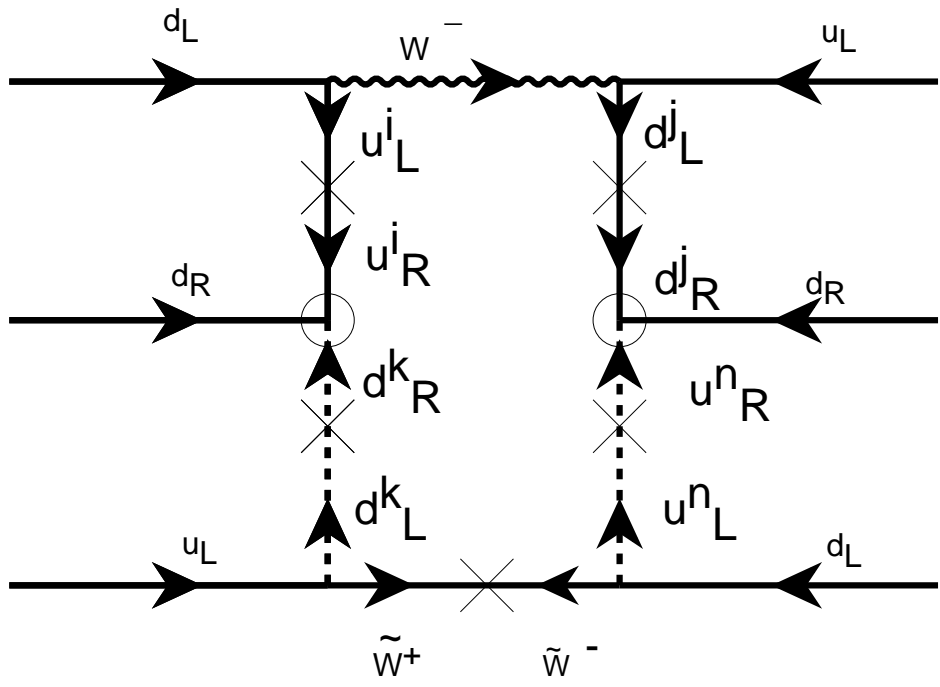
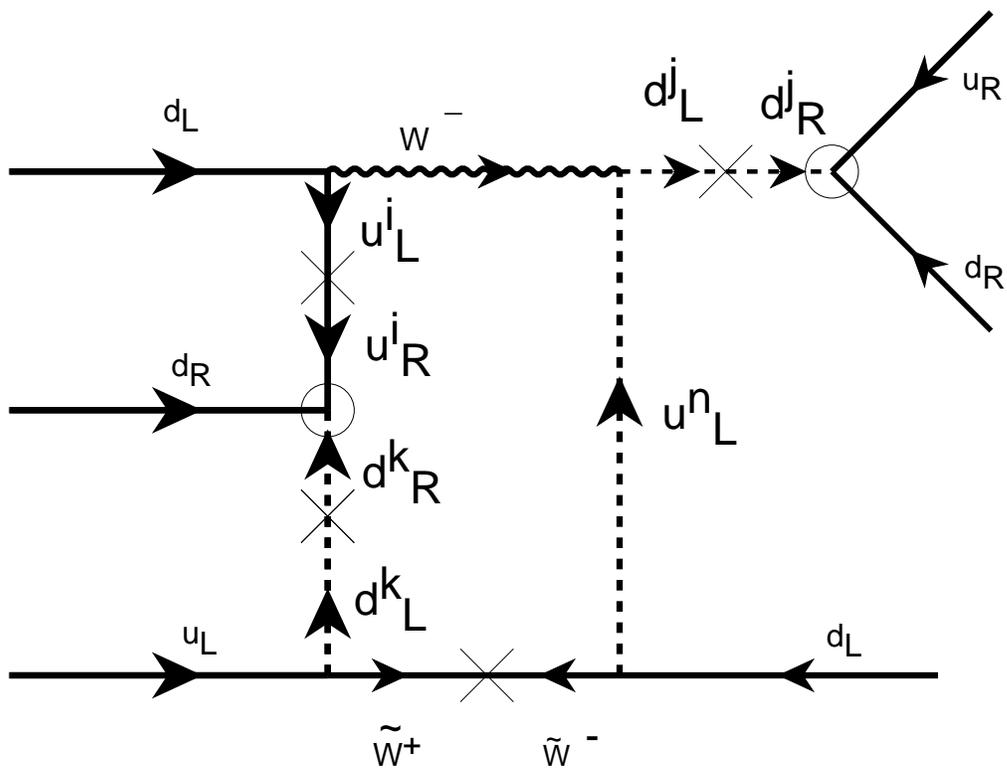
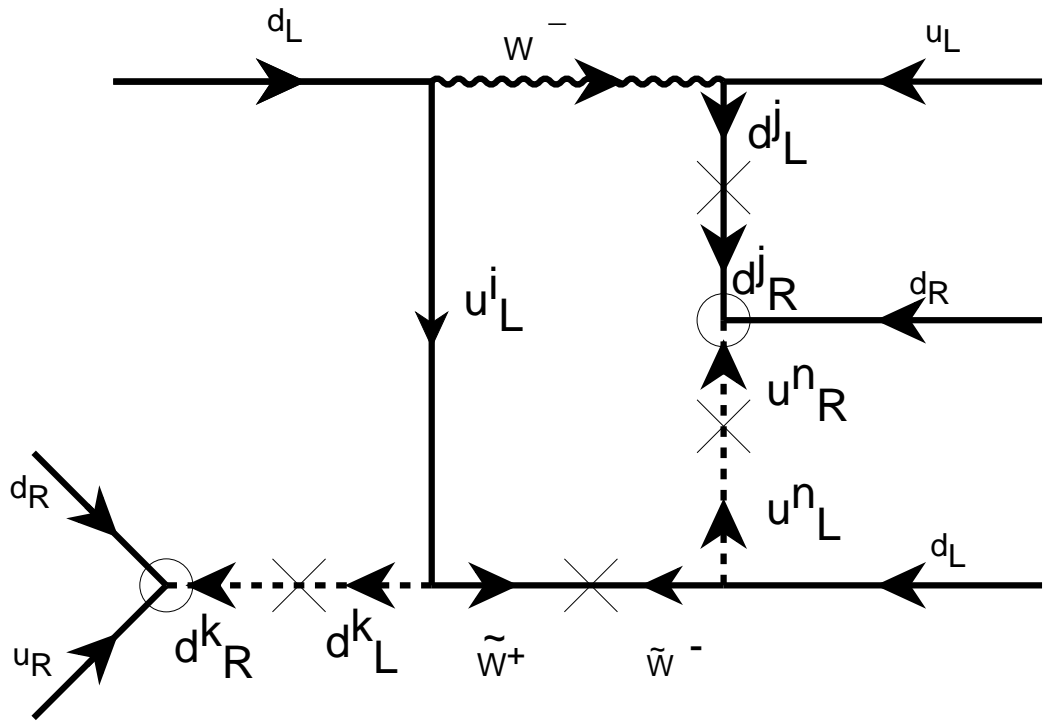
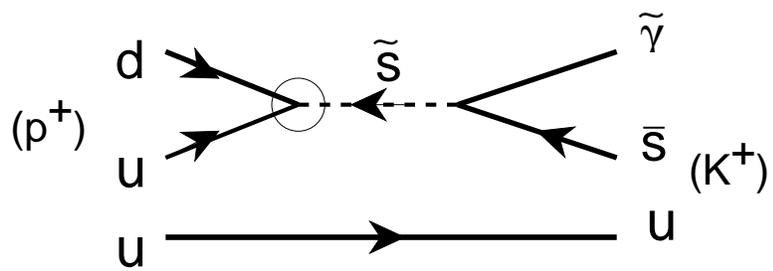
\end{document}